\documentclass[twocolumn,aps,pre,showpacs,preprintnumbers,amsmath,amssymb,floatfix]{revtex4}

\usepackage{graphicx}
\usepackage[usenames,dvipsnames]{color}

\newlength{\onefig}
\newlength{\twofig}
\newcommand{\ve}[1]{\mathbf{#1}}

\begin{document}

\title{Hopping and microscopic dynamics of ultrasoft particles in cluster crystals}

\author{Daniele Coslovich}
\email[Corresponding author: ]{daniele.coslovich@univ-montp2.fr}
\author{Lukas Strauss}
\thanks{D. Coslovich and L. Strauss contributed equally to this work.}
\author{Gerhard Kahl}
\affiliation{Institut f\"ur Theoretische Physik and Center for
  Computational Materials Science (CMS), Technische Universit\"at
  Wien, Wiedner Hauptstra{\ss}e 8-10, A-1040 Wien, Austria}

\date{\today}

\begin{abstract}
  We have investigated the slow dynamics of ultrasoft particles in
  crystalline cluster phases, where point particles interact through
  the generalized exponential potential $u(r) = \epsilon
  \exp[-(r/\sigma)^n]$, focusing on the cluster fcc phase of this
  model with $n=4$. In an effort to elucidate how the mechanisms of
  mass transport depend on the microscopic dynamics and in order to
  mimic a realistic scenario in a related experiment we have performed
  molecular dynamics, Brownian dynamics, and Monte Carlo simulations.
  In molecular dynamics simulations the diffusion of particles
  proceeds through long-range jumps, guided by strong correlations in
  the momentum direction. In Monte Carlo and Brownian dynamics
  simulations jump events are short-ranged, reflecting the purely
  configurational nature of the dynamics. In contrast to what was
  found in models of glass-forming liquids, the effect of Newtonian
  and stochastic microscopic dynamics on the long-time relaxation
  cannot be accounted for by a temperature-independent rescaling of
  the time units. From the obvious qualitative discrepancies in the
  short time behavior between the three simulation methods and the
  non-trivial difference in jump length distributions, long time
  relaxation, and dynamic heterogeneity, we learn that a more complex
  modeling of the dynamics in realistic systems of ultrasoft colloids is required.
\end{abstract}

\pacs{61.20.Ja, 64.70.pv, 82.70.Dd}

\maketitle

\section{Introduction}
\label{sec:introduction}

Although the formation of stable clusters of particles that interact
via entirely repulsive, ultrasoft potentials might seem
counterintuitive at first sight, it is by now well established that
cluster phases of such systems \textit{do}
exist~\cite{klein_repulsive_1994,likos_2001,mladek_formation_2006,mladek_erratum:_2006,likos_do_2007,mladek_multiple_2008}.
This phenomenon is potentially relevant for a wide class of ultrasoft
colloidal systems, such as dendrimers and microgels, in which the
centers of mass of the colloidal particles can overlap with a finite
energy cost. Theoretical considerations~\cite{mladek_clustering_2007}
in combination with computer simulations~\cite{mladek_phase_2007} for
models of ultrasoft colloidal particles have given evidence for the
existence of these clusters and have, in addition, provided a deeper
understanding of this phenomenon. Stable clusters form either in
disordered, liquid-like phases or occur as ordered cluster crystals
where the particle aggregates populate the lattice sites of regular
fcc or bcc lattices. Most of the investigations have been dedicated up
to now to study the static properties of the cluster phases, of which
we mention the two most remarkable ones: (i) freezing and melting
lines depend in a linear way on the density and (ii) the lattice
constant of cluster crystals is invariant under compression, inducing
thereby a linear growth of the cluster size with density.

In contrast, little effort has been dedicated, so far, to obtain a
deeper insight into the dynamics that governs these cluster forming
systems.  To the best of our knowledge, only two contributions that
deal with the diffusion and the relaxation dynamics in cluster
crystals have been
published~\cite{moreno_diffusion_2007,likos_cluster-forming_2008}. From
these investigations, based on molecular dynamics (MD) simulations, it
became clear that the dynamics in cluster crystals shows features at
least as intriguing as their static counterparts. For instance, these investigations
revealed that particles move constantly between
neighboring clusters, while maintaining the original lattice structure
of the cluster crystal and the average cluster population on the
lattice sites. Further, a closer analysis of the dynamic correlation
functions provided evidence for a decoupling between self and
collective time-dependent correlations~\cite{moreno_diffusion_2007}.

The present contribution is dedicated to investigate the dynamical
properties of ultrasoft particles in a crystalline cluster phase in
detail and from a more general point of view. In particular, we intend
to study how mass transport is realized in different simulation
schemes, which mimic different scenarios in related realistic systems:
(i) in MD simulations, where the influence of the microscopic solvent
is neglected and the mesoscopic (colloidal) particles move according
to Newton's law; (ii) in Monte Carlo (MC) simulations, where the random
collisions of the colloids with the solvent particles are mimicked
through the stochastic nature of the algorithm; (iii) in Brownian
dynamics (BD) simulations, where the influence of the solvent is taken into
account via the friction term and the stochastic forces acting on the
particles.

The model system that we choose to study cluster phases is the generalized
exponential model of index $n$ (GEM-$n$)~\cite{mladek_formation_2006},
in which particles interact via the potential
\begin{equation}
\label{eqn:gem}
\Phi(r) = \epsilon \exp[-(r/\sigma)^n] \,.
\end{equation}
In the literature, this model has been used to describe the effective
interactions between linear polymer chains
($n=2$)~\cite{louis_can_2000} and dendrimers ($n\approx
3$)~\cite{mladek_computer_2008} in the dilute regime. For $n>2$, we
expect this model to capture the general features of purely repulsive,
cluster-forming systems~\cite{likos_do_2007}. In contrast to
Ref.~\cite{moreno_diffusion_2007}, where $n=8$ was considered, we have
based our investigations on the GEM-4 system, for which highly
accurate data for the phase diagram are available
\cite{mladek_formation_2006,mladek_phase_2007}. In a first step, we
have focused on the microscopic dynamics and on the jump events,
comparing MD and MC dynamics.  We have found distinct differences
between these two types of dynamics imposed by the simulation
technique: in MC simulations, hopping takes place only over short
distances (i.e., essentially from one cluster to a neighboring one);
in contrast, in MD simulations particles can migrate over large
distances through the crystal.  Furthermore, our investigations on the
diffusion constant and on the dynamic correlation functions have shown
that properties calculated from MD and MC {\it cannot} be superposed
by a simple, temperature-independent rescaling of time. 
This finding is in
striking contrast to what is known from
normal~\cite{huitema_can_1999,rutkai_dynamic_2010} or
glass-forming~\cite{berthier_monte_2007,gleim_relaxation_1998,berthier_revisitingslow_2007}
liquids. Finally, we have found that BD simulations give rise to
dynamical properties very similar to those observed in MC
simulations, supporting the view that MC dynamics effectively
(although approximately) incorporates solvent
effects~\cite{kikuchi_1991,brambilla_probingequilibrium_2009,sanz_marenduzzo_2010}.
Our results indicate that a complex dynamical scenario may arise as a
competition of activated processes and momentum correlations, and
suggest that the solvent properties may change qualitatively the nature of
the slow dynamics in cluster-forming colloidal systems.

The paper is organized as follows: in Section~\ref{sec:model} we
introduce the model and present our simulation methods; in
Section~\ref{sec:results} we present our results for the dynamical
properties of a GEM-4 cluster crystal and finally in
Section~\ref{sec:conclusions} we give our conclusions. In the Appendix
we present our algorithm for cluster identification.

\section{Model and methods}
\label{sec:model}

Our model is composed of $N$ classical particles of mass $m$ enclosed
in a cubic box with periodic boundary conditions. Particles interact
through the generalized exponential potential, Eq.~\eqref{eqn:gem},
with $n=4$. The potential is truncated and shifted at
$r_c=2.2\sigma$. In the following, we will use the parameters $\sigma$
and $\epsilon$ in Eq.~\eqref{eqn:gem} as units of distance and energy,
respectively. The units of time are chosen differently according to
the simulation method (see below). For each studied density in the fcc
cluster phase (see Fig.~\ref{fig:phase_diagram}), the
\textit{equilibrium} value of $N$ was determined using the algorithm
developed in Ref.~\cite{mladek_phase_2007}, yielding $N=3367$ for
$\rho=6.4$. During this procedure, we used an fcc cluster crystal
comprising four unit cells per side. We have checked that size effects do
not play a relevant role by performing simulations on smaller systems,
i.e., using three unit cells per side.

The dynamical properties of the model were investigated using MD,
MC, and BD simulations. MD simulations were performed in the $NVE$
ensemble with fixed center of mass using the velocity-Verlet
algorithm~\cite{allen_computer_1987}. Equilibration at various
temperatures was achieved by reselecting the velocities of particles
at regular time intervals according to the appropriate Maxwellian
distribution. MC simulations were performed in the $NVT$ ensemble
using the standard Metropolis algorithm. Attempted moves involved only
single particle displacements, randomly generated over a cube of side
0.6. The acceptance ratio ranged from 64 \% (at high $T$) to 42 \% (at low $T$). 
At any fixed $T$, the acceptance ratio decreases
monotonically with the length of the maximum attempted displacement.
Thus, in contrast to
what found for a model glass-forming
liquid~\cite{berthier_monte_2007}, it is not possible to define an
``optimal'' value of the maximum attempted displacement. The value
chosen in this work is a reasonable comprise between efficiency of the
simulation and physical realism. BD simulations were performed by
integrating a set of coupled, inertia-free Langevin equations using a
simple Euler algorithm~\cite{allen_computer_1987}. We used a friction
parameter $\xi=10^{-3}$ (in reduced units) and a temperature-dependent time step $\delta
t_\text{BD}$. $\xi$ was chosen sufficiently small that the dynamical
properties at high $T$ no longer depend on $\xi$. To avoid artifacts in the dynamic correlation
functions measured during MC and BD simulations, the motion of the
center of mass of the system was subtracted out from the particle
displacements. In the following, we will use
$\sqrt{m\sigma^2/\epsilon}$ as the unit of time for MD simulations. In
these units, the time step $\delta t_\text{MD}$ for the integration of the
equations of motion was 0.03. In the case of MC simulations, time will
be measured in units of MC sweeps, where one MC sweep corresponds to
$N$ attempted displacement moves. In BD simulations, the time unit
will be $\sigma^2/D_0$, where $D_0=1/\xi$ is the diffusion coefficient at
infinite dilution and unit temperature. $\delta
t_\text{BD}$ was adjusted at each state point so that the dynamical
properties did not depend on it anymore upon further reduction, and
ranged from $5 \cdot 10^{-6}$ (at low $T$) to $10^{-8}$ (at high $T$). Procedures to match the time
scales of the different simulations will be discussed in
Sec.~\ref{sec:results}.

\section{Results}
\label{sec:results}

\begin{figure}[t]
\includegraphics[width=\onefig]{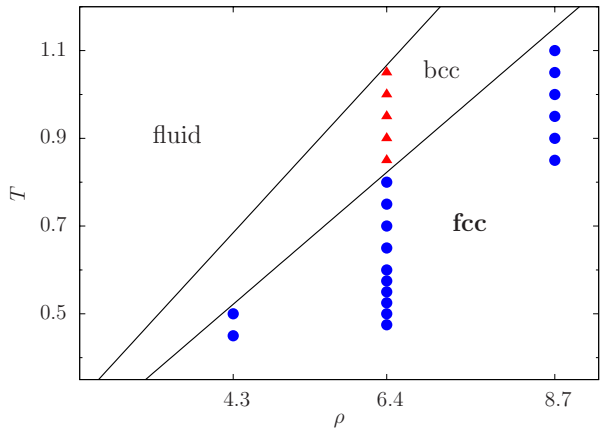}
\caption{\label{fig:phase_diagram} State points studied in this work
  in the temperature-density phase diagram. Triangles
  indicate state points for which the fcc cluster crystal was
  metastable. 
  The lines between cluster fcc,
  cluster bcc, and fluid phases of the GEM-4 system have been redrawn
  from Ref.~\cite{mladek_formation_2006}.}
\end{figure}

In this section we characterize the dynamical properties of GEM-4
particles in the fcc cluster crystal phase, paying particular
attention to jump diffusion processes and to the effects of the 
different types of microscopic dynamics introduced in
Sec.~\ref{sec:model}. We will focus on the behavior of the system as a
function of temperature along the isochore $\rho=6.4$ (see
Fig.~\ref{fig:phase_diagram}). If not stated otherwise, in the
following we will always refer to this particular density. The
investigated temperature range comprises both the stable fcc cluster
phase ($T\alt 0.80$) and the super-heated regime. In the latter,
the underlying fcc crystal structure remained (meta)stable throughout
the simulation. We also performed simulations for selected state
points at different densities ($\rho=4.3$ and 8.7, see
Fig.~\ref{fig:phase_diagram}) and checked that the expected scaling of
dynamic quantities as a function of $\rho/T$
holds~\cite{moreno_diffusion_2007}. In the following we will therefore
use $\rho/T$ as the natural control parameter of the system. In
Sec.~\ref{sec:jumps} and ~\ref{sec:diffusion} we will focus our
attention on the comparison between MD and MC simulations. In
Sec.~\ref{sec:bd} we will discuss the results of BD
simulations.

\subsection{Microscopic dynamics and jump events}\label{sec:jumps}

\begin{figure}[t]
\includegraphics[width=\onefig]{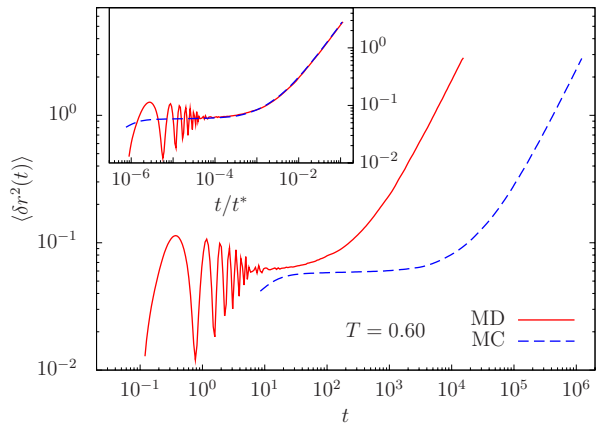}
\caption{\label{fig:msd} Main panel: mean square displacement $\langle
  \delta r^2(t)\rangle$ as a function of $t$ at $\rho=6.4$
  and $T=0.60$ from MD (solid line) and MC (dashed line)
  simulations. Inset: same as main panel but as a function of the
  rescaled time $t/t^*$ (see text for
  definition).}
\end{figure}

\begin{figure}[t]
\includegraphics[width=\onefig]{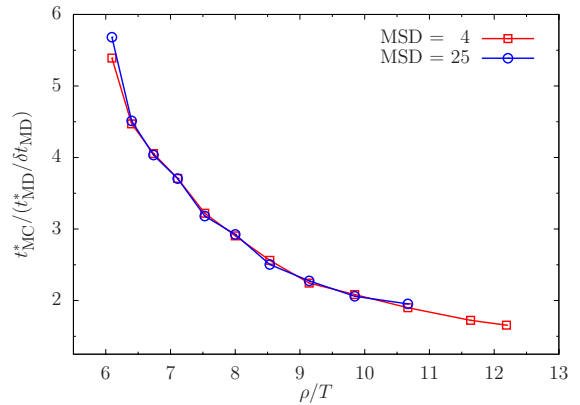}
\caption{\label{fig:taumc} Time scaling factor
  $t^*_\text{MC}/(t^*_\text{MD}/\delta t_\text{MD})$ as a function of
  $\rho/T$. $\delta t_\text{MD}$ is the time step used in MD
  simulations. $t^*_\text{MC}$ and $t^*_\text{MD}$ are defined as the
  times at which the mean square displacement equals a fixed target
  value in MC and MD simulations, respectively. The target values
  considered are 4 (squares) or 25 (circles).}
\end{figure}

We start our discussion with the analysis of the mean square
displacement $\langle\delta r^2(t)\rangle = \langle |\ve{r}(t) -
\ve{r}(0)|^2 \rangle$. This allows us to review some general features
of the dynamics discussed previously for GEM-8
particles~\cite{moreno_diffusion_2007} and to perform a first
comparison of MD and MC simulations. In the main panel of
Fig.~\ref{fig:msd} we show the mean square displacement for a
representative state point well within the stable fcc phase
($\rho=6.4$, $T=0.60$). 
After the ballistic time range (not shown here), the MD dataset shows
distinct oscillations due to single-particle vibrational
modes~\cite{likos_do_2007,moreno_diffusion_2007,strauss_diploma},
which disappear for $t \agt 10$. At long times, normal
diffusion is recovered, i.e., $\langle\delta r^2(t)\rangle \sim
6Dt$. Due to the arbitrary choice of the ``time unit'' in the case of
MC simulations, the respective curve is shifted. It is possible,
however, to rescale the time variable, $t/t^*$, so that the long time
behavior of $\langle \delta r^2(t/t^*)\rangle$ coincides for both
types of dynamics. The state-dependent scaling times $t^*$ are
defined, for both dynamics, by the condition $\langle\delta
r^2(t^*)\rangle = 25$. The precise choice of the ``target'' mean square
displacement is irrelevant as long as this value is within the
diffusive regime (see also Fig.~\ref{fig:taumc}). The original and the
rescaled datasets are shown in the main panel and inset of
Fig.~\ref{fig:msd}, respectively. Adopting the rescaled time
representation, the mean square displacements obtained from MD and MC
dynamics nicely collapse onto a unique curve at long times
($t/t^* \agt 10^{-4}$). Clear discrepancies are found at short
times, due to the expected phonon suppression in MC
dynamics~\cite{berthier_monte_2007}. Similar rescaling approaches were
used in previous works on Lennard-Jones
liquids~\cite{huitema_can_1999,berthier_monte_2007}. However, in
contrast to what was found in those studies, the ratio of the MD and
MC scaling times displays a systematic temperature dependence. This is
demonstrated in Fig.~\ref{fig:taumc}, where we show the time scaling
factor $t^*_\text{MC}/(t^*_\text{MD}/\delta t_\text{MD})$ for two values of the
target mean square displacement. The ratio
$t^*_\text{MC}/(t^*_\text{MD}/\delta t_\text{MD})$ decreases monotonically with
$1/T$ within the explored temperature range. Therefore, it is not
possible to set, once and for all, the time conversion factor between
the different types of dynamics.

To understand the discrepancy between the thermal behaviors of the
system in MD and MC simulations, we analyze in more detail both types
of dynamics and identify the elementary hopping processes leading to
particle diffusion. To this end, we monitor individual particle
trajectories and identify jump events by means of the cluster analysis
outlined in the Appendix. Our definition of jump events is as
follows. At any instant of time $t$, we map the position $\ve{r}(t)$
of a tagged particle to the center of mass (CM) of its respective,
initial cluster, located at
$\ve{R}_\text{initial}^\text{cm}=\ve{R}^\text{cm}(t)$. In MD
simulations, a particle is considered ``equilibrated'' in its
respective cluster if the residence time exceeds a characteristic
time, $t^*_\text{eq}$, of the order of a few vibrational periods
(cf. Fig.~\ref{fig:msd}). In our calculations at $\rho=6.4$, we chose
$t^*_\text{eq}=3.6$, independent of temperature. We checked that
reasonable variations of $t^*_\text{eq}$ around the selected value by
some 10\% or 20\%, or introduction of a $T$-dependent $t^*_\text{eq}$
do not modify qualitatively our analysis. In MC simulations, the
equilibration time is given by $t^*_\text{eq}
(t^*_\text{MC}/t^*_\text{MD})$, where $t^*_\text{MC}$ and
$t^*_\text{MD}$ are the scaling times introduced above. A jump event
starts when the tagged particle leaves the cluster in which it was
originally
equilibrated 
and ends when the residence time of the tagged particle
in any of the visited clusters exceeds $t^*_\text{eq}$. The \emph{net}
jump vector is then $\ve{r}_\text{net}=\ve{R}^\text{cm}_\text{final} - \ve{R}^\text{cm}_\text{initial}$, where
$\ve{R}^\text{cm}_\text{final}$ is the CM position of the final cluster site visited
during the jump process. This analysis is
carried out for all jumping particles in the system and averages are
performed over all identified jump events.

\begin{figure}[t]
\includegraphics[width=\onefig]{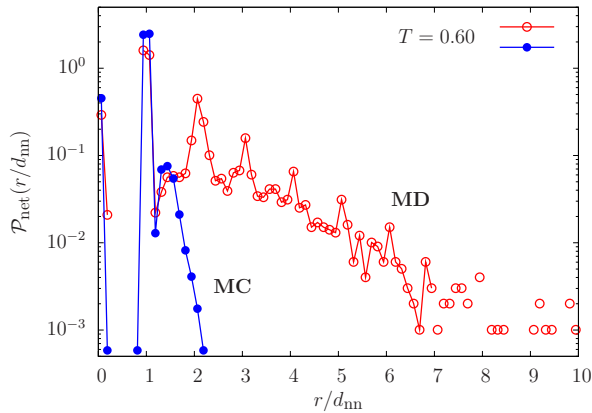}
\caption{\label{fig:jumplength_mcmd} Distribution of net jump length
  $\mathcal{P}_\text{net}$ as a function of $r/d_\text{nn}$ at
  $\rho=6.4$ and $T=0.60$ for MD (empty circles) and MC (filled circles)
  simulations.}
\end{figure}

The normalized distribution of net jump lengths
$\mathcal{P}_\text{net}(r) = \mathcal{P}(|\ve{r}_\text{net}|)$ is
shown in Fig.~\ref{fig:jumplength_mcmd} for the same state point as
considered in Fig.~\ref{fig:msd}. The MD results display marked peaks
corresponding to integer multiples of the nearest neighbor positions in
the fcc lattice, suggesting preferential motion along straight
paths. Strikingly, the distribution extends up to and even beyond
$10d_\text{nn}$, where $d_\text{nn}$ is the nearest neighbor
distance. That is, particle diffusion in MD simulations proceeds
through correlated, long range jumps, whose typical length exceeds
$d_\text{nn}$. The estimated probability of long-range jumps,
$\int_{d_\text{nn}}^{\infty} dr \mathcal{P}_\text{net}(r)$, is
approximately 50\% at the temperature considered in
Fig.~\ref{fig:jumplength_mcmd}, while the estimated probability of
jumping to the nearest neighbor site is $\sim 30\%$. These values do
not depend significantly on temperature (see below). Hence, correlated
long-range jumps provide the leading mechanism for mass transport
during MD simulations. The shape of the distribution of net jump
lengths changes considerably when the MC dynamics is considered. In
this case, $\mathcal{P}_\text{net}(r)$ decays rapidly beyond $r\sim
d_\text{nn}$. This can be understood in terms of the stochastic nature
of the MC dynamics and suggests that long range jumps in MD
simulations are guided by correlations in the momentum direction along
the jump path. We will discuss this point in more detail below.

\begin{figure}[t]
\includegraphics[width=\onefig]{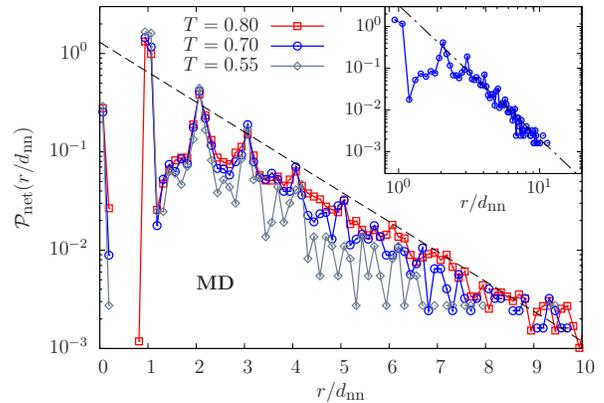}
\caption{\label{fig:jumplength_md} Main panel: distribution of net
  jump length $\mathcal{P}_\text{net}$ as a function of
  $r/d_\text{nn}$ for MD simulations at various temperatures. The
  dashed line indicates an exponential function $\sim \exp(-r/\bar{r})$,
  with $\bar{r}\approx 1.43 d_{\text{nn}}$. Inset: log-log plot of
  $\mathcal{P}_\text{net}(r/d_\text{nn})$ for $T=0.70$. The
  dash-dotted line is a power law decay $1/r^{1+\alpha}$ with $\alpha=2.2
  \pm 0.2$.}
\end{figure}

For both types of microscopic dynamics, the net jump length
distribution $\mathcal{P}_\text{net}(r)$ displays a mild dependence on
temperature. In Fig.~\ref{fig:jumplength_md} we show
$\mathcal{P}_\text{net}(r)$ for MD simulations at temperatures $T = 0.80$, 0.70, and
0.55 for MD simulations. This temperature range corresponds to a variation of diffusivity
of approximately three decades (see Fig.~\ref{fig:arrhenius}, further
discussed below). In general, $\mathcal{P}_\text{net}(r)$
decays a bit more rapidly as $T$ decreases. However, the probability of long
range jumps in MD simulations remains substantial even at the lowest
investigated temperature ($T=0.55$). Hence, the qualitative
differences between MD and MC simulations discussed above with reference to Fig.~\ref{fig:jumplength_mcmd} are relevant
for the whole investigated $T$ range. At large distances ($r>2
d_\text{nn}$) the envelope of the maxima of
$\mathcal{P}_\text{net}(r)$ obtained from MD simulations can be
described reasonably well by an exponential function,
$a \exp{(-r/{\bar{r}})}$, over approximately two decades in
$\mathcal{P}_\text{net}(r)$. $\bar{r}$ is a characteristic jump length, which depends
very weakly on $T$ and whose value is $\approx 1.43 d_\text{nn}$. 

As it can be seen from the inset of Fig.~\ref{fig:jumplength_md}, at
even larger distances the shape of $\mathcal{P}_\text{net}(r)$ seems
to crossover to a power law decay, $1/r^{1+\alpha}$, a behavior
reminiscent of Levy
flights~\cite{shlesinger_strange_1993,zaslavsky_chaos_2002}. The
estimated power law exponent ($1+\alpha \approx 3.2$) is slightly
higher than the upper bound required for standard Levy
flights~\cite{shlesinger_strange_1993,zaslavsky_chaos_2002}. Furthermore,
inspection of the mean square displacement (see Fig.~\ref{fig:msd})
does \textit{not} show clear signatures of anomalous diffusion---a
feature commonly associated with Levy
flights~\cite{shlesinger_strange_1993}---on any time scale accessible
by our MD simulations. On the basis of the current data, it remains
unclear whether a Levy flights picture would be appropriate for the
description of diffusion in cluster crystals of ultrasoft
particles. More extensive investigations are required to completely
settle this issue.

\begin{figure}
\includegraphics[width=\onefig]{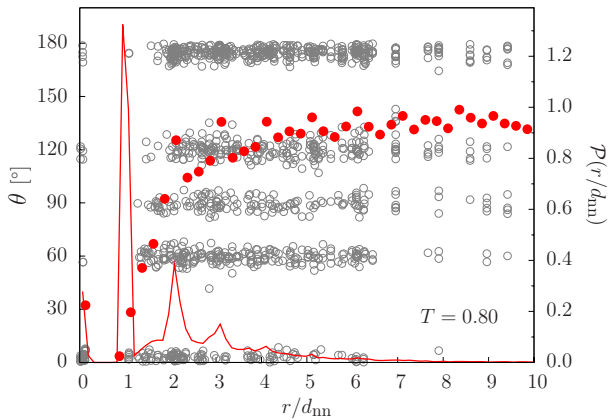}
\caption{\label{fig:angle_jumplength} Correlation between angles
  $\theta$ and net jump lengths $r$ along jump paths at $T=0.80$ from
  MD simulations. Empty circles: angles $\theta$ between successive
  jumps versus net jump length $r$. Filled circles: correlation
  between average values of angles at fixed jump length and the jump
  lengths. The solid line indicates the corresponding distribution of
  net jump lengths $\mathcal{P}_\text{net}(r/d_\text{nn})$
  (cf. Fig.~\ref{fig:jumplength_md}).}
\end{figure}

\begin{figure}
\includegraphics[width=0.7\onefig]{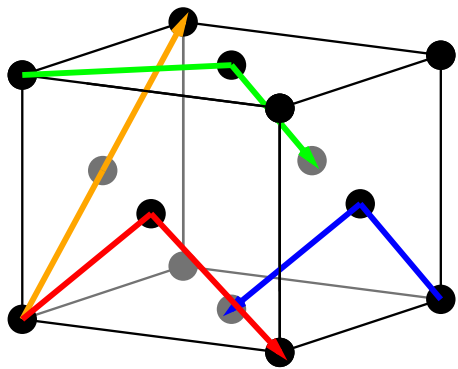}
\caption{\label{fig:directions} Possible consecutive jumps to nearest
  neighbors in the fcc unit cell. The full black and grey circles
  indicate fcc lattice sites. Blue arrow: $60^{\circ}$,
  $r=d_\text{nn}$.  Red arrow: $90^{\circ}$, $r=\sqrt{2}d_\text{nn}$.
  Green arrow: $120^{\circ}$, $r=\sqrt{3}d_\text{nn}$.  Orange arrow:
  $180^{\circ}$, $r=2d_\text{nn}$.  }
\end{figure}

\begin{figure*}
\includegraphics*[width=0.90\textwidth]{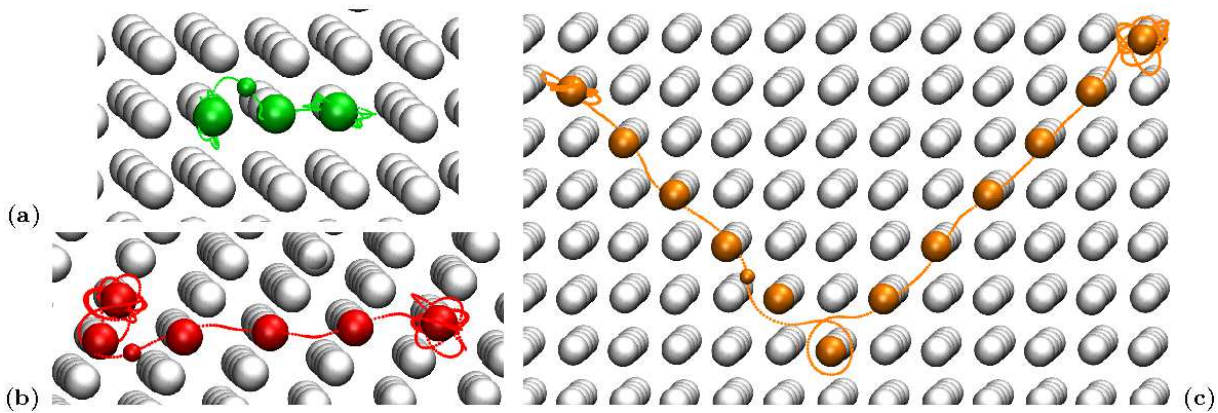}
\caption{\label{fig:snapshots} Representative jump events of tagged
  particles (indicated by the colored small interstitial sphere)
  during MD simulations at $T=0.80$. Large particles indicate the
  centers of mass of the visited cluster sites. The cluster sites
  visited during the jump path are highlighted with large colored
  spheres. The positions of the tagged particle along the trajectory
  are marked at equidistant steps.}
\end{figure*}

To better characterize the nature of the long range jumps occurring in
MD simulations, we now study the relationship between the net jump
length $r$ and the angles $\theta$ enclosed by successive steps
between cluster sites along the jump path. In
Fig.~\ref{fig:angle_jumplength} we display the measured angles
$\theta$ versus the corresponding \textit{net} travelled distance $r$
for a representative state point ($T=0.80$). For each jump of length
$r$ we stored $n-1$ angles, where $n$ was the number of steps along the
path. The horizontal bands at $60^{\circ}$, $90^{\circ}$,
$120^{\circ}$, and $180^{\circ}$ reflect the possible angles between
consecutive steps of length $d_\text{nn}$ in the fcc lattice (see
Fig.~\ref{fig:directions}). Angles $\theta\approx 0$ occur for consecutive
back and forth steps, while the data at $r\approx 0$ are associated to
jumps, possibly comprising several intermediate steps, whose initial
and final clusters coincide.

Fig.~\ref{fig:angle_jumplength} shows that the data tend to
accumulate at large angles, $\theta=180^{\circ}$ and $120^{\circ}$,
corresponding to jumps to second nearest neighbors of the fcc lattice
along straight, or slightly deflected, directions. The filled circles in
Fig.~\ref{fig:angle_jumplength} indicate the average angles at a fixed
average net jump length. On average, large angles correspond to large
jump lengths, revealing that particles tend to proceed along
approximately straight paths, possibly keeping a significant
correlation in the direction of their momentum along the jump path.
To illustrate the peculiar nature of the hopping processes observed in
MD simulations, we show in Fig.~\ref{fig:snapshots} the trajectories
of selected particles at $T=0.80$. They display correlated jumps over
different lengths, ranging from a few nearest neighbor distances
(Fig.~\ref{fig:snapshots}-a) to over $\sim 10$ nearest neighbors
distances (Fig.~\ref{fig:snapshots}-c). This latter event clearly
shows a strong directional correlation of the movement, persisting
over several intermediate steps along the jump path.

\subsection{Diffusion, heterogeneity, and relaxation}\label{sec:diffusion}

\begin{figure}[t]
\includegraphics[width=\onefig]{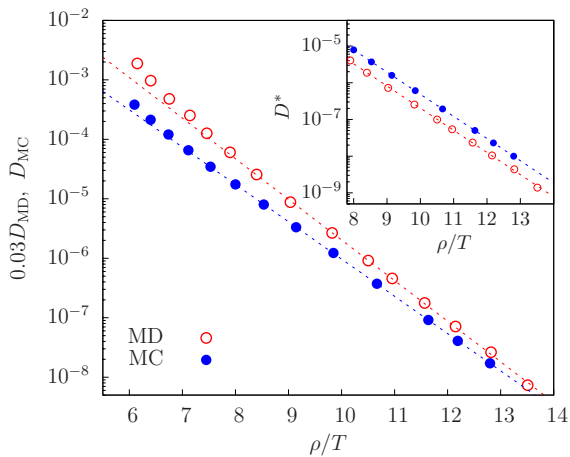}
\caption{\label{fig:arrhenius} Main panel: Arrhenius plot of the
  diffusion coefficient $D_\text{MD}$ from MD (empty symbols) and
  $D_\text{MC}$ from MC (filled symbols) simulations. Inset: scaled
  diffusion coefficients $D^*$ obtained from MD and MC simulations
  (see text for definition). In both panels, dotted lines represent
  fits to the Arrhenius law $D_0 \exp(-E/T)$, where $E$ is an
  effective activation energy. The activation energy $E$ equals
  $1.57\rho$ and $1.44\rho$ for fits to $D_\text{MD}(T)$ and
  $D_\text{MC}(T)$, respectively, and $1.39\rho$ for fits to $D^*(T)$,
  independent of the microscopic dynamics.}
\end{figure}

Naturally, the question arises: how much are transport properties
affected by the difference in the microscopic dynamics? To discuss
this point, we compare in Fig.~\ref{fig:arrhenius} the $T$-dependence
of the diffusion coefficient $D$ obtained from MD and MC simulations
using the standard Einstein relation, i.e., $\langle\delta
r^2(t)\rangle = 6Dt$. As it is customary for systems displaying slow
dynamics, the Arrhenius representation is adopted. In the main panel
of Fig.~\ref{fig:arrhenius} the diffusion coefficient for MD,
$D_\text{MD}$, is multiplied by 0.03 for better comparison with the MC
data.
The fact that it is not possible to present both sets of data on a
single curve using a simple rescaling of the time units, demonstrated
in the figure by the different slopes of the two data sets, is
consistent with the systematic $T$-dependence of the ratio of scaling
times $t^*_\text{MC}/t^*_\text{MD}$ (cf. Fig.~\ref{fig:taumc}). This
finding is particularly relevant in view of previous numerical
investigations on liquids in
equilibrium~\cite{huitema_can_1999,rutkai_dynamic_2010} or in the
supercooled
regime~\cite{gleim_relaxation_1998,berthier_monte_2007,berthier_revisitingslow_2007}.
In the latter three studies, MD and Brownian
dynamics~\cite{gleim_relaxation_1998} or MD and MC
dynamics~\cite{berthier_monte_2007,berthier_revisitingslow_2007} were
used. It was shown that the long time relaxation of glass-forming
liquids is \textit{independent} of the microscopic dynamics. This
means that relaxation times, transport coefficients or dynamic
correlation functions obtained via the different simulation methods
can be superposed after a $T$-independent rescaling of the time
units. For the slow dynamics of ultrasoft particles in cluster
crystals, however, such a rescaling approach is hindered due to the
prominent role of momentum correlations. Hence, our results indicate
that independence of long time relaxation from the microscopic
dynamics solely holds for systems where the relevant dynamical
correlations are \textit{configurational} in nature.

MD simulations of a GEM-8 cluster crystal~\cite{moreno_diffusion_2007}
have shown that the $T$-dependence of $D$ closely follows the
Arrhenius law, $D = D_0\exp{(-E/T)}$, where $E=E(\rho)$ is an
effective activation energy that scales proportionally to
$\rho$~\cite{moreno_diffusion_2007,strauss_diploma}. This observation,
combined with our analysis of the jump length distributions
(Figs.~\ref{fig:jumplength_mcmd} and~\ref{fig:jumplength_md}),
suggests a way to rationalize the discrepancy between MC and MD
results for $D$. This can be achieved by invoking a thermally
activated diffusion process governed by a unique activation energy,
i.e., independent of the microscopic dynamics, but based on different
distributions of elementary jump lengths. Towards this aim, let us
write down the Arrhenius law for activated diffusion assuming a
distribution $\mathcal{P}_\text{net}(r)$ of jump lengths
\begin{equation}
  \label{eq:arr}
  D = \frac{\int_0^\infty \!\! r^2 \mathcal{P}_\text{net}(r) dr}{\langle\tau_w\rangle} \exp{(-E/T)} , 
\end{equation}
where $\langle\tau_w\rangle$ is an average waiting time, which
ultimately sets the natural time scale for the process. As long as the
$T$-dependence of $\langle \tau_w \rangle$ is approximately similar
for the different dynamics, it should be possible to rectify the
Arrhenius plots of Fig.~\ref{fig:arrhenius} by considering the
rescaled diffusion coefficient $D^* = D / \int_0^\infty \!\!  r^2
\mathcal{P}_\text{net}(r) dr$. The inset of Fig.~\ref{fig:arrhenius}
shows that this is indeed the case.
Fits to an Arrhenius law for $D^*(T)$ provide a unique effective
activation energy $E\approx 1.39\rho$, independent of microscopic
dynamics. We found that the fitted value of $E$ is close to the
barrier height $E_b$ separating neighboring clusters,
estimated~\cite{strauss_diploma} from single particle potential
energies ($E_b\approx 0.94 E$). This clearly points to a
non-cooperative nature of activated dynamics in the system under
consideration.

\begin{figure}
\includegraphics[width=\onefig]{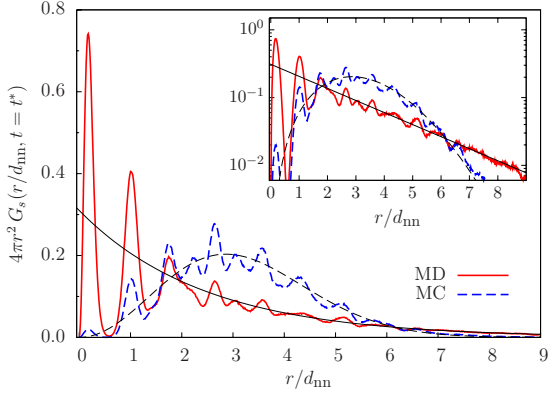}
\caption{\label{fig:vanhove} Main panel: Self part of the van Hove
  correlation function $G_s(r,t=t^*)$ as a function of $r/d_\text{nn}$
  at $T=0.80$ for MD (thick solid line) and MC (thick dashed line)
  simulations. The time $t^*$ is chosen such that $\langle\delta
  r^2(t^*)\rangle=25$. The thin dashed line indicates the expected
  functional form for standard random walk, $4\pi {r^2}
  \exp[-r^2/(4Dt)]/(4\pi Dt)^{3/2}$. The thin solid line indicates
  an exponential function $\sim \exp(-r/\bar{r})$, {with $\bar{r}\approx 2.44
    d_\text{nn}$}. Inset: same as main plot but on a semi-logarithmic scale.}
\end{figure}

The microscopic dynamics exerts even a stronger influence on
other typical time-dependent correlation functions of the liquid
state, such as density-density time correlators. In
Fig.~\ref{fig:vanhove} we show the self part of the van Hove
correlation function $G_s(r,t)$ at $T=0.80$ and compare MD and MC
datasets for $t=t^*$, where again the scaling time $t^*$ is chosen
such that $\langle\delta r^2(t^*)\rangle=25$. The MC dataset can be
simply described by a Gaussian function with variance $2Dt$ (as
expected for standard random walks) modulated by peaks centered at the
nearest neighbor distances of the fcc crystal. The MD dataset, on the
other hand, is much more stretched and displays an approximately
exponential envelope. This behavior is a direct consequence of the
broad distribution of elementary jump lengths
(cf. Fig.~\ref{fig:jumplength_mcmd} and~\ref{fig:jumplength_md}).

\begin{figure}[t]
\includegraphics[width=\onefig]{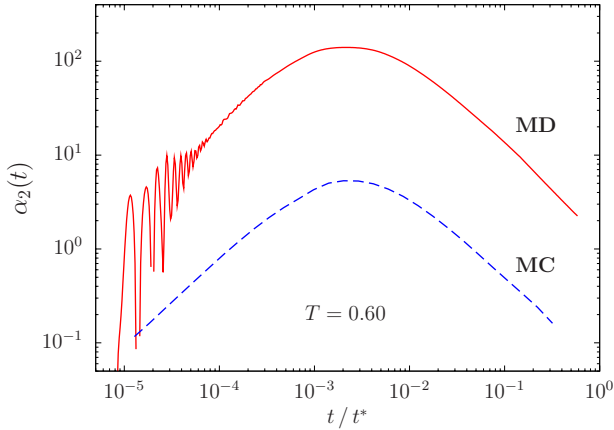}
\caption{\label{fig:nongaussian} Non-Gaussian parameter $\alpha_2$ as
  defined in Eq.~\eqref{eqn:alpha2} as a function of the rescaled time
  $t/t^*$ at $T=0.60$ from MD (solid line) and MC (dashed line)
  simulations with $t^*$ as defined in the text.}
\end{figure}

\begin{figure}
\includegraphics[width=\onefig]{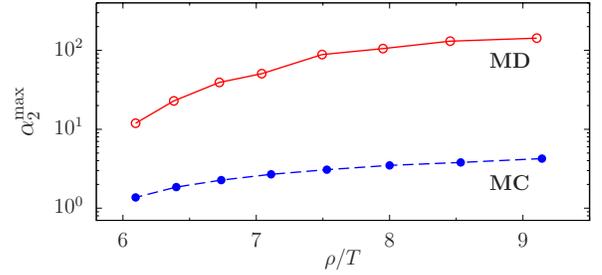}
\caption{\label{fig:nongaussian_tmax} Maximum value of the
  non-Gaussian parameter $\alpha_2^\text{max}$ as a function of
  $\rho/T$ for MD (empty circles) and MC (filled circles) simulations.}
\end{figure}

The stretched shape of the van Hove correlation function obtained from
MD simulations motivates a closer inspection of the non-Gaussian
parameter
\begin{equation}
\label{eqn:alpha2}
\alpha_2(t)=\frac{3\langle \delta r^4(t)\rangle}{5{\langle
    \delta r^2(t)\rangle}^2}-1 \, ,
\end{equation}
which quantifies the deviation from Gaussianity of the distribution of
particle displacements at time $t$. In the context of glass-forming
liquids, $\alpha_2(t)$ has also been
used~\cite{vogel_temperature_2004,coslovich_dynamics_2009} as a simple
measure of ``dynamic
heterogeneity''~\cite{donati_stringlike_1998,widmer-cooper_reproducible_2004,appignanesi_democratic_2006}. In
Fig.~\ref{fig:nongaussian} we show $\alpha_2(t/t^*)$ at $T=0.60$.
At short times
qualitative discrepancies appear between MD and MC data, as expected:
marked oscillations in $\alpha_2(t/t^*)$, due to single-particle
vibrations, are observed in fact only for MD simulations. At longer
times, on the other hand, the overall shape of $\alpha_2(t/t^*)$ is
very similar to the one found in normal and supercooled
liquids~\cite{coslovich_dynamics_2009}: for both types of simulations,
$\alpha_2$ displays a peak at intermediate times corresponding to a
maximum value $\alpha_2^\text{max}$. As in standard glass-formers,
$\alpha_2^\text{max}$ grows by decreasing $T$, albeit in a moderate
fashion (see Fig.~\ref{fig:nongaussian_tmax}). However, the values of
$\alpha_2^\text{max}$ obtained in MD simulations are significantly
larger (up to a factor $40$) than in the MC simulations of our system
or than the typical values observed in MD simulations of model
glass-forming liquids~\cite{coslovich_dynamics_2009}. This is a
consequence of the broad distribution of elementary jump lengths found
in MD simulations for the system under consideration. Such a dynamic
heterogeneity is essentially non-cooperative in origin and therefore
different from the one normally observed in glass-forming
liquids~\cite{donati_stringlike_1998,widmer-cooper_reproducible_2004,appignanesi_democratic_2006}. By
visual inspection of animated trajectories, we have checked that jump
events are not associated to correlated motions of several particles
(as observed, for instance, in~\cite{donati_stringlike_1998}), but
rather reflect the migration of individual particles over the sample.

\begin{figure*}[t]
\includegraphics*[width=\twofig]{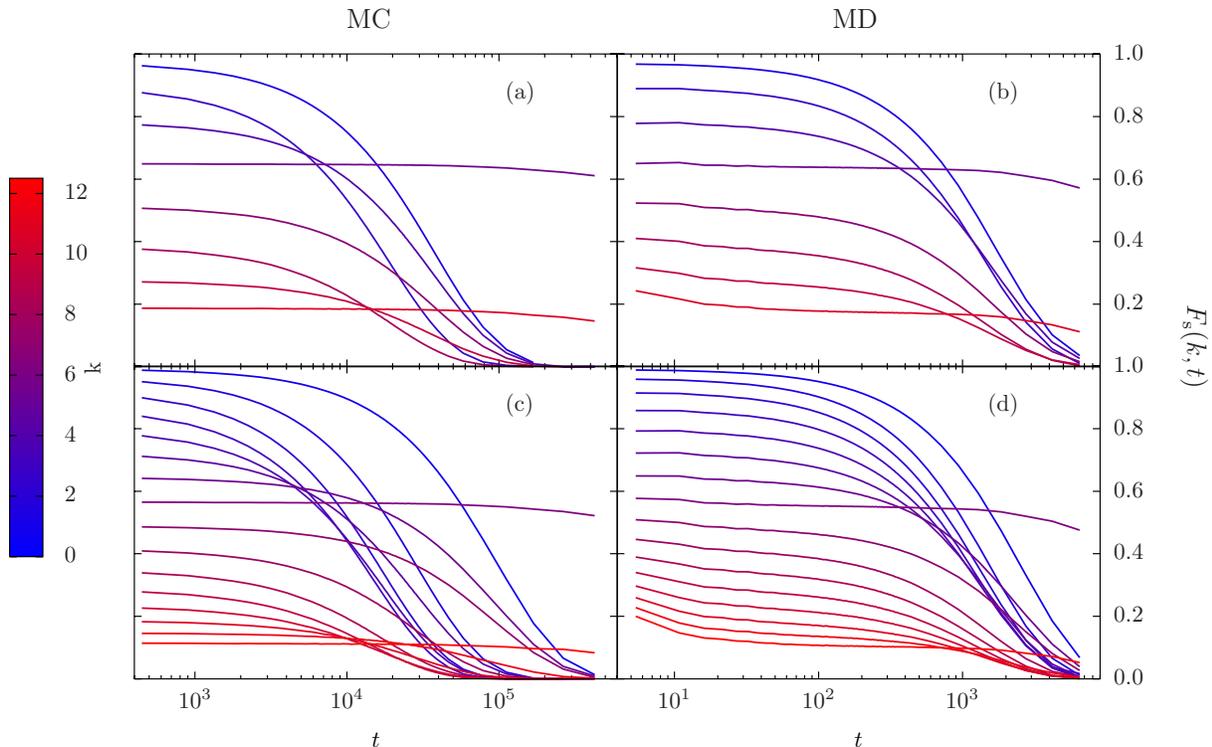}
\caption{\label{fig:fkt} Self intermediate scattering function
  $F_s(\ve{k},t)$ as a function of $t$ at $T=0.80$. The wave-vectors
  considered are chosen along the directions $\ve{k}_1$ (panels (a)
  and (b)) and $\ve{k}_2$ (panels (c) and (d)) corresponding to the
  first and second reciprocal lattice wave-vectors of the fcc lattice,
  respectively. The norm of $\ve{k}$ is varied from 0 to $2 k_i$, with
  $i=1,2$ (curves from bottom to top); the curves are colored
  according to the colorscale shown on the left side of the
  figure. Left and right panels are results from MC and MD
  simulations, respectively.}
\end{figure*}

Finally, we study the dependence of the self intermediate scattering
function
$$
F_s(\ve{k},t) =
\frac{1}{N}\sum_{j=1}^{N} \big\langle \exp
\left\{i \ve{k} \cdot [\ve{r}_j(t)-\ve{r}_j(0)]\right\} \big\rangle
$$
on the microscopic dynamics. Due to the underlying crystal structure,
the relaxation of $F_s(\ve{k},t)$ is strongly anisotropic, i.e., it
depends explicitly on the direction of $\ve{k}$. While this effect of
anisotropy is expected to be most pronounced at reciprocal lattice
wave-vectors, we found in addition that $F_s(\ve{k},t)$ possesses a
non-trivial angular dependence at \textit{any} fixed norm (not shown
here). To simplify the discussion, in the following we focus only on
directions in reciprocal space corresponding to the first and second
reciprocal lattice wave-vectors of the fcc lattice, $\ve{k}_1/{k}_1$
and $\ve{k}_2/{k}_2$. Along each of these directions, we vary the norm
of $\ve{k}$ within the range $0 < k \leq 2 k_{i}$, where $k_{i}$ is
the norm of the selected reciprocal lattice vector (${k}_1=5.39$,
${k}_2=6.23$). Results obtained at $T=0.80$ are collected in
Fig.~\ref{fig:fkt}. Clearly, the relaxation of $F_s(\ve{k},t)$ becomes
extremely slow at reciprocal lattice vectors. 
However, the relaxation patterns for off-lattice
wave-vectors are rather complex in the two dynamics: as the norm of
$\ve{k}$ approaches that of a reciprocal wave-vector the relaxation
slows down, but the difference in relaxation times between ``fast''
and ``slow'' wave-vectors depends markedly on the microscopic
dynamics. Our results indicate a larger heterogeneity of relaxation
times for MC dynamics, a fact which contrasts our previous analysis
of the non-Gaussian parameter. A similar conclusion is drawn from the
analysis of the angular dependence of $F_s(\ve{k},t)$ at fixed norm of
$\ve{k}$ (not shown here). Further investigations are thus required to
clarify this behavior and its possible connection to the decoupling
between self and collective relaxation discussed
in Ref.~\cite{moreno_diffusion_2007}.

\subsection{Comparison with Brownian dynamics}\label{sec:bd}

Our analysis has focused so far on the comparison of dynamical
properties obtained through MD and MC simulations. MC simulations are
expected to mimic in some sense the Brownian motion of colloidal
particles due to collisions with solvent particles. Simulation data~\cite{kikuchi_1991} and approximate analytical
models~\cite{sanz_marenduzzo_2010} provide support for this view, at
least for small MC displacements. In this subsection we explicitly
compare BD and MC simulations for the GEM-4 model, and show that the
two methods lead indeed to very similar macroscopic dynamics.

\begin{figure}
\includegraphics[width=\onefig]{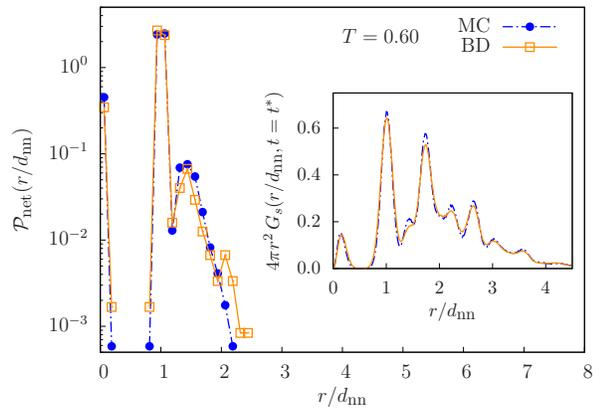}
\caption{\label{fig:bd1} Main panel: Distribution of net jump length
  $\mathcal{P}_\text{net}$ as a function of $r/d_\text{nn}$ at
  $\rho=6.4$ and $T=0.60$ for BD (empty squares) and MC (filled
  circles) simulations. Inset: Self part of the van Hove
  correlation function $G_s(r,t=t^*)$ as a function of $r/d_\text{nn}$
  at $T=0.60$ for BD (solid line) and MC (dash-dotted line)
  simulations. The time $t^*$ is chosen such that $\langle\delta
  r^2(t^*)\rangle=9$.}
\end{figure}

We start with a direct assessment of the net jump length distribution
$\mathcal{P}_\text{net}(r/d_\text{nn})$ obtained from BD simulations at
$T=0.60$. The results are displayed in the main panel of
Fig.~\ref{fig:bd1}, and compared with the corresponding dataset from MC
simulations already presented in Fig.~\ref{fig:jumplength_mcmd}. The net jump length distribution from BD simulations is
short-ranged and remarkably similar to that obtained using MC
simulations. A small peak at $r\approx 2d_\text{nn}$ indicates
rare jumps over two nearest neighbor distances, a feature
absent in the MC dataset. In the inset of Fig.~\ref{fig:bd1} we show
the van Hove correlation function $G_s(r,t^*)$ calculated at a time
$t^*$ at which the MSD equals 9. We observe a striking correspondence
between BD and MC datasets, demonstrating a close similarity between
the dynamics emerging from the two simulation methods.

\begin{figure}
\includegraphics[width=\onefig]{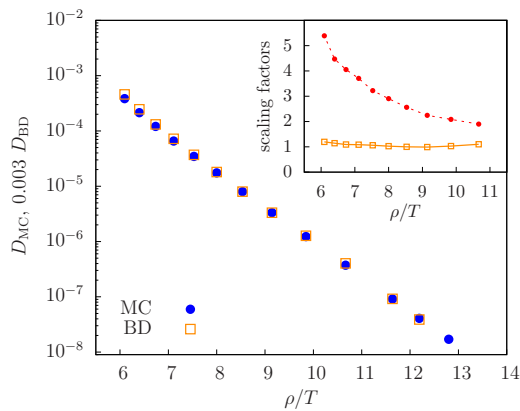}
\caption{\label{fig:bd2} Main panel: Arrhenius plot of the diffusion
  coefficient $0.003\cdot D_\text{BD}$ from BD (empty squares) and
  $D_\text{MC}$ from MC (filled circles) simulations. The scaling
  factor for $D_\text{BD}$ has been chosen so as to maximize the
  overlap of the two data sets. Inset: time scaling factors $t^*_\text{MC}/(t^*_\text{BD}/0.003)$ (open squares) and $t^*_\text{MC}/(t^*_\text{MD}/\delta t_\text{MD})$ (filled circles) as a
  function of $\rho/T$. $t^*_\text{MC}$, $t^*_\text{MD}$, and $t^*_\text{BD}$ are
  defined as the times at which the mean square displacement equals 4
  in MC, MD, and BD simulations, respectively.}
\end{figure}

The temperature dependence of the diffusion coefficient $D(T)$ is
analyzed in Fig.~\ref{fig:bd2} employing the Arrhenius
representation. Within the temperature regime explored by BD
simulations, the $T$-dependences of the diffusion coefficients
obtained from BD and MC simulations closely follow one another, up to
a trivial, state-independent rescaling of the time unit. Thus, not only
the qualitative aspects of transport mechanisms, but also the
$T$-dependences of transport coefficients generated by BD and MC are
very similar. To contrast the results obtained through these
stochastic methods to those of Newtonian dynamics, we report in
the inset of Fig.~\ref{fig:bd2} the ratio of the scaling times
$t^*_\text{MC}/t^*_\text{BD}$ and $t^*_\text{MC}/t^*_\text{MD}$, as defined in
Sec.~\ref{sec:jumps}. As expected, $t^*_\text{MC}/t^*_\text{BD}$ remains fairly
constant throughout the available temperature regime. Within the same
$T$-range, $t^*_\text{MC}/t^*_\text{MD}$ displays a systematic variation,
as already evidenced in the main panel of Fig.~\ref{fig:bd2}.

Overall, the results shown above indicate a strong similarity between
the transport mechanisms at play in BD and MC simulations and
corroborate the analysis outlined in the previous subsections. In this
respect, the results shown in Fig.~\ref{fig:bd1} and \ref{fig:bd2} support
the use of MC simulations as an effective way to incorporate solvent
effects in a computer simulation of model
colloids~\cite{brambilla_probingequilibrium_2009,sanz_marenduzzo_2010}. We
remark that an attempt to rescale the time unit in our MC
simulations by the acceptance ratio, as suggested
in~\cite{sanz_marenduzzo_2010}, resulted in slightly poorer agreement
between the two datasets of $D(T)$. Thus, more extensive
investigations on the connection between dynamical properties generated
by MC and BD simulations and possible system-specific aspects are
required.

\section{Conclusions and outlook}
\label{sec:conclusions}

We have performed detailed MD, MC, and BD simulations for ultrasoft
particles that form stable clusters of overlapping particles, which,
in turn, populate the lattice sites of a regular fcc lattice. Our
investigations on the dynamical properties of the system (in terms of
particle moves, diffusion and dynamic correlation functions)
revealed striking differences between Newtonian and stochastic
simulation methods.

(i) In MD simulations, particle diffusion is realized as long-ranged
sequences of jumps from one cluster to a neighboring one, assisted by
strong correlations in momentum directions. The envelope of the jump
length distribution is within a good approximation an exponentially
decaying function of the distance and extends up to and beyond ten
nearest neighbor distances. At large distances, the jump length
distribution is described rather well by a power law,
$1/r^{1+\alpha}$, with $\alpha\approx 2.2$. In MC simulations, by
contrast, hopping events are limited to the nearest clusters, as a
result of the purely configurational nature of the dynamics. (ii) The
diffusion constant, $D$, evaluated in MD and MC simulations cannot be
matched on a single master curve by a simple rescaling of the
respective time units. This finding is in striking contrast to related
results obtained in investigations on
equilibrated~\cite{huitema_can_1999,rutkai_dynamic_2010} and
supercooled~\cite{gleim_relaxation_1998,berthier_monte_2007,berthier_revisitingslow_2007}
liquids. However, by normalizing $D$ with respect to the underlying
jump length distribution, an activation energy can be filtered out
that is common to both types of dynamics. (iii) Also for the dynamic
correlation functions we find distinct differences, which become
apparent in the overall shape of the self part of the van Hove
correlation function, in the non-Gaussian parameter, and in the long
time decay of the self intermediate scattering function for
off-lattice wave vectors. (iv) BD simulations give rise to dynamical
properties very similar to those observed during MC simulations,
supporting the view that solvent effects are implicitly and
effectively included in MC simulations through the stochastic nature
of the algorithm.

The present contribution---along with
Refs.~\cite{moreno_diffusion_2007,likos_cluster-forming_2008}---can
only be considered as a first step towards a deeper understanding of
the evidently complex dynamics of cluster forming, ultrasoft
systems. Quite a few issues remain unaddressed, which will be dealt
with in future contributions. First, a more systematic study of the
influence of the solvent, using full Langevin dynamics and possibly
including hydrodynamic interactions, is required and will be performed
in future work. Other simulation methods, such as dissipative particle
dynamics~\cite{hoogerbrugge_simulating_1992,nikunen_would_2003}, might
provide a different, complementary view on our data. Another aspect is
related to the underlying model: our inter-particle potential
represents an effective interaction between two mesoscopic (colloidal)
particles, where the degrees of freedom of the microscopic
constituents of the colloids have been traced
out~\cite{likos_effective_2001}. An important issue is thus related to
the \textit{level} of coarse-graining, ranging from the present
picture of an ``effective''
particle~\cite{likos_effective_2001,mladek_computer_2008}, over
partially averaging over sub-entities, such as in multi-blob
representations~\cite{pierleoni_soft_2007}, or force-matching
schemes~\cite{izvekov_multiscale_2005,noid_multiscale_2008}, to the
fully atomistic level where all the constituent entities of the
colloidal particles and the particles of the solvent are considered
explicitly. Another fundamental question addressing the interpretation
of the dynamic correlation functions is how the dynamics of a system
of ``effective'' particles is related to that of a system of
atomistically resolved particles.

In conclusion, our work has revealed qualitative differences in the
dynamics of ultrasoft, cluster-forming particles obtained using
Newtonian and two stochastic (i.e., Brownian dynamics and Monte Carlo) simulation methods. This effect has been
attributed to the competition between activated processes and solvent
effects, which controls the distribution of jump lengths and
ultimately the dynamic correlation functions. We speculate that the
interplay between activated slow dynamics and solvent effects could
generate complex dynamical scenarios in ultrasoft colloids, such as
dendrimers and microgels, and even a broader class of cluster-forming
colloids, including systems with competing interactions. With this contribution we hope to motivate further numerical and experimental studies on the dynamics of such colloidal systems to test this hypothesis.

\begin{acknowledgements}
  We thank B. M. Mladek, D. Frenkel, C.~N. Likos, W. Kob, F. Colonna
  for useful comments and discussions. We also thank C.~N. Likos for
  a critical reading of the manuscript. Financial support by the
  Austrian Science Fund (FWF) under project P19890-N16 and the Center
  for Computational Materials Science (CMS) is acknowledged.
\end{acknowledgements}

\appendix

\section{Cluster identification algorithm}
\label{sec:appendix}

\begin{figure}[t]
\includegraphics[width=\onefig]{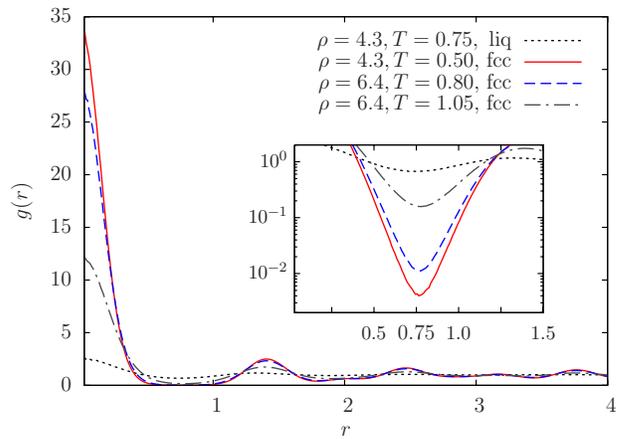}
\caption{\label{fig:gr_cluster} Radial distribution function, $g(r)$,
  of a GEM-4 system as a function of $r$ for different densities
  $\rho$ and temperatures $T$, both in the liquid and fcc cluster
  crystal phase. From the inset we see that $g(r)$ does not vanish
  completely at its first minimum, $r_\text{min} \approx 0.75$.}
\end{figure}

In an effort to trace the migration of the particles through the
cluster crystal we developed an algorithm to
distinguish between different clusters and to identify the affiliation
of a particular particle to a cluster in an unambiguous way at every
step of the simulation. In previous investigations on clustering
systems~\cite{mladek_clustering_2007}, the position of the first
minimum in the pair distribution function $g(r)$, $r_{\rm min}$, was used to provide information
whether two particles, separated by a distance $r$, belong to the same
cluster (for $r \le r_{\rm min}$) or not (for $r > r_{\rm min}$). This
criterion can undoubtedly be used to obtain a first, rough estimate
for identifying those particles that belong to a particular cluster.
However, a closer analysis of $g(r)$ reveals
(cf. Fig.~\ref{fig:gr_cluster}) that even for relatively low
temperatures this function does not vanish completely for $r \sim
r_{\rm min}$. Particles migrating between two neighboring cluster
sites of the cluster crystals are to be made responsible for this effect.

\begin{figure}[t]
\includegraphics[width=\onefig]{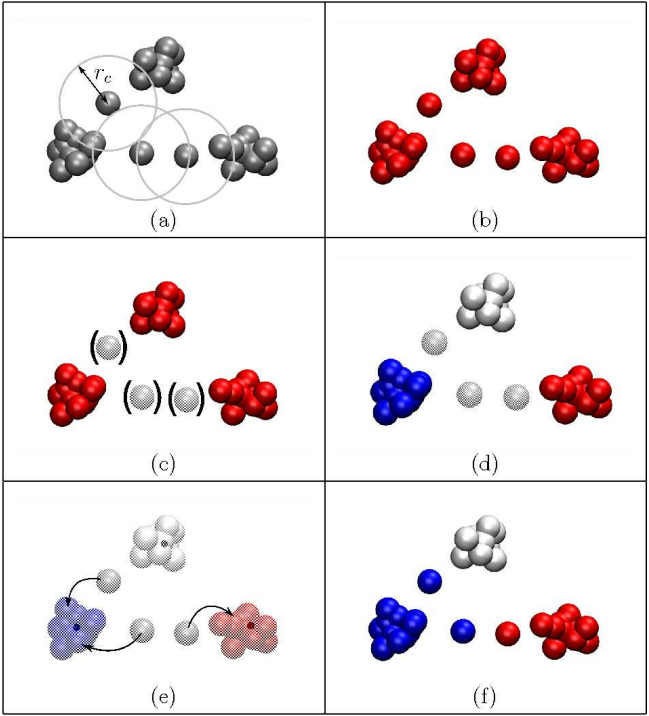}
\caption{\label{fig:cluster} Procedure to separate merged
  clusters. (a) Three neighboring clusters are merged by hopping
  particles. At this step, all particles in red form a single
  cluster. (b) The search for particle neighbors is repeated for the
  merged cluster with a reduced cut-off radius $r_c$. (c) Particles
  (light grey spheres) with a small number of neighbors
  ($n_c<n_c^\text{min}$) are excluded. (d) The search for particle
  neighbors is repeated ignoring the light grey spheres, identifying
  thereby disjointed clusters. (e) Excluded particles (light grey spheres)
  are reintroduced and reassigned to the respective nearest
  clusters. (f) Procedure accomplished, yielding three disjoint
  clusters.}
\end{figure}

In the following we present a refined version of the cluster
identification algorithm presented in Ref.~\cite{mladek_clustering_2007};
it can be subdivided into the following four steps:

\begin{figure}[t]
\includegraphics[width=\onefig]{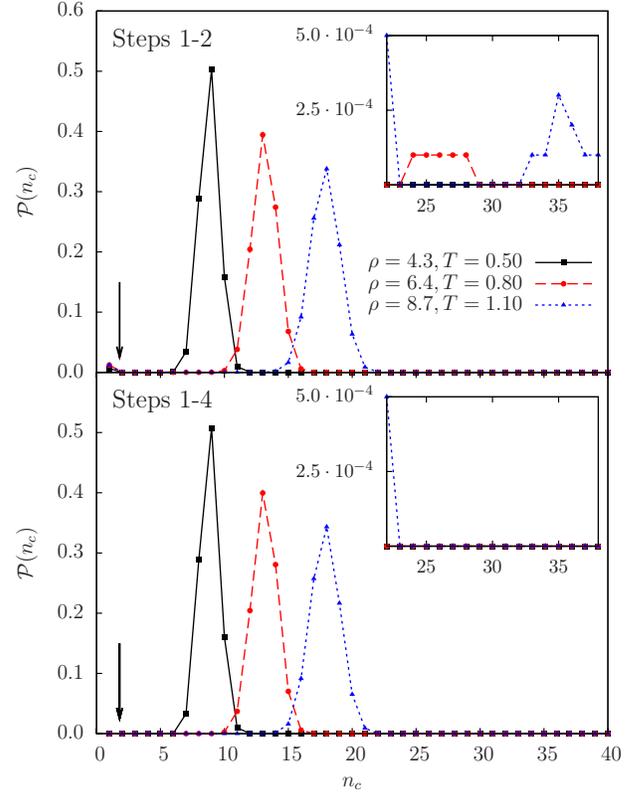}
\caption{\label{fig:nc_cluster} Distribution of cluster sizes $n_c$
  for various state points. Upper panel: distribution calculated after
  Step 2 of the cluster analysis algorithm. The inset shows an enlargement of
  the large-$n_c$ region, revealing the occurrence of merged
  clusters. The arrow indicates the occurrence of clusters composed
  by few particles. Lower panel: distribution calculated after Step 4. The
  inset reveals the absence of merged clusters. The arrow indicates
  that all clusters of size less than $n_c^\text{min}$ have vanished.}
\end{figure}

\begin{enumerate}
\item We start our procedure with the first particle and identify all
  particles separated by a distance less than a given
  cut-off radius $r_c$ as neighboring particles. This procedure is
  repeated for all remaining particles. As a first guess we choose
  $r_c = r_{\rm min}$, a value which will be corrected iteratively
  during this algorithm.
\item With this information at hand, we start again with the first
  particle and label all its neighbors, their respective neighbors,
  etc.; in this way, we are able to identify a first cluster of the
  system. We then proceed to the next particle that has not been
  labelled yet, and repeat the operation. Finally, all particles 
  have been assigned to a cluster.
\item At this stage, the algorithm reproduces exactly the results
  obtained in~\cite{mladek_clustering_2007}. It risks, however, to
  provide misleading data: as particles move from one cluster to
  another, the particles of these two clusters might now be counted to
  belong to the set of neighbors of the hopping particle, merging
  thereby the two clusters (see Fig.~\ref{fig:cluster}-a). In our refined
  algorithm, we modify the first guess of $r_c$ in an iterative way
  and eventually all merged clusters and all single-particle
  clusters are eliminated. To this end, we reduce $r_c$
  and introduce three check parameters: $n_c^{\rm min}$ and $n_c^{\rm
    max}$, the expected minimum and maximum cluster size present
  in our system (roughly estimated from the cluster size
  distribution calculated after Steps 1 and 2) and $N_c$, the number
  of lattice sites in our system which is, by definition, equal to the
  amount of clusters in the crystal phase.

  We proceed as follows: all clusters that have just been identified
  are re-considered. If the size of one of them exceeds $n_c^{\rm
    max}$, the corresponding collection of particles is isolated from
  the rest of the system (see Fig.~\ref{fig:cluster}-a) and the search
  for neighbors is repeated within the remaining set of particles
  (Fig.~\ref{fig:cluster}-b). In a similar manner, those particles
  with the lowest number of neighbors are also isolated from the
  others (Fig.~\ref{fig:cluster}-c) since these particles are found to
  be responsible for merging neighboring clusters. Ignoring the excluded particles for a moment, the neighbors are identified once
  again as described above (cf. Step 2), giving disjoint clusters
  (Fig~\ref{fig:cluster}-d). Now the isolated particles are
  re-integrated into the ensemble and are reassigned to those
  clusters with the nearest center of mass position
  (Fig.~\ref{fig:cluster}-e and \ref{fig:cluster}-f). Finally, the
  following checks are made: (i) Does the size of all newly
  established clusters lie within $n_c^{\rm max}$ and $n_c^{\rm min}$?
  (ii) Is the number of clusters identified in the system equal to the
  number of lattice sites, $N_c$? If one of these two conditions is
  violated, the procedure is iterated reducing the cut-off radius $r_c$
  at each iteration step.

\item In a final check on the size of the clusters, those particles or
  small collections of particles that possibly have not been assigned
  to none of the clusters before, are assigned to the cluster with
  the nearest center of mass. The success of the improved cluster
  analysis algorithm becomes obvious from the ensuing cluster
  population distribution. Peaks due to single particles and merged
  clusters have vanished (see Fig.~\ref{fig:nc_cluster}, right panel),
  reflecting the correct analysis of the cluster sizes distribution of
  the system.
\end{enumerate}

Finally, to match the identity of clusters at different times, we
exploit the fact that in the \textit{crystal phase} the centers of mass of
the clusters are fixed to their respective lattice sites. To be more
specific, we found that the root mean square displacement of the
centers of mass does not exceed ten percent of the nearest-neighbor
distance between clusters, $d_\text{nn}$, at any state point in the fcc
region of the phase diagram. We can thus keep track of
the identity of any selected cluster by locating the cluster that, at
the next time step, has the closest center of mass and by repeating
this operation for all time steps.


\end{document}